\documentclass[3p,twocolumn]{elsarticle}

\usepackage{lineno,hyperref}
\modulolinenumbers[5]

\journal{Journal of Molecular Liquids}

\begin{document}

\begin{frontmatter}

\title{A Network Approach to Unravel Correlated Ion Pair Dynamics in Protic Ionic Liquids. The Case of Triethylammonium Nitrate}
\author{Tobias Zentel}
\ead{tobias.zentel@uni-rostock.de}
\author{Oliver K\"uhn\corref{cor1}}
\ead{oliver.kuehn@uni-rostock.de}
\address{Institute  of Physics, University of Rostock, Albert-Einstein-Str. 23-24, 18059 Rostock, Germany }
\cortext[cor1]{Corresponding author}
\date{today}
\begin{abstract}
The intermolecular interactions in the title compound are investigated using self-consistent charge density functional based tight binding molecular dynamics. Emphasis is put on the analysis of correlated motions of ion pairs using  ideas of network theory. At equilibrium such correlations are not very pronounced on average. However, there exist sizeable local  correlations for cases where  two cations share the same anion via two NHO-hydrogen bonds. The effect of an external perturbation, which artificially introduces a sudden local heating of an NH-bond, is investigated using nonequilibrium molecular dynamics. Here, it is found that the average N-H bond vibrational relaxation time is about 5.3 ~ps. This energy redistribution is rather nonspecific with respect to the ion pairs and does not lead to long-range correlations spreading from the initially excited ion pair.
\end{abstract}


\end{frontmatter}


\section{Introduction}
Ionic liquids (ILs) are promising candidates for a wide variety of applications, most importantly as 'green' solvents \cite{rogers03_792}. To exploit the exceptional physico-chemical properties in novel applications, the influence of intermolecular interactions must be understood in great detail. The ionic character of the molecules leads to strong Coulomb interactions, but there are also dispersion forces \cite{Ludwig15_13790} and a wide range of Hydrogen bond (HB) interactions of different strength  \cite{hunt15_1257}. The HBs may link ions of opposite charge into networks. Here,  protic ILs are of particular interest due to their rather strong HBs. Further, these systems allow for  proton transfer from the cation to the anion such as to create a neutral pair.  

The strong Coulomb interactions lead to a long range ordering, which can be seen, for example, in the radial distribution functions \cite{delpopolo04_1744,Maginn09_373101}. Furthermore, the polar and nonpolar groups can segregate and form spatial heterogeneities in the liquid \cite{Wang05_12192}. For example, Canongia Lopes and co-workers  analysed the mesoscopic segregation of imidazolium-based ILs depending on the alkyl chain lengths \cite{Shimizu14_567,Bernardes14_6885}.  However, the substructure of the polar regions is still  not solved and the concept of ion pairing and long-lived neutral subunits is much debated \cite{Kirchner15_463002}. Important in this context is the charge transfer between anion and cation, leading to absolute ion charges below one $e$ \cite{Holloczki14_16880}. Zhang and Maginn showed that the  ion pair lifetimes can be linked to transport properties \cite{Zhang15_700}. The effects of noncovalent intermolecular forces can be quantified by quantum chemical simulations on small systems down to the atomic resolution \cite{Marekha15_16846}. Experimentally, these interactions can be probed by NMR chemical shift measurements; see, for example, the study of the effect of mixtures on HB interactions in Ref.~\cite{Marekha15_23183}. 
 The intermolecular interactions give rise to signatures in infrared absorption and Raman spectra \cite{Roth12_105026}. Both dispersion and HB interactions show spectral signs in the far infrared region  \cite{fumino09_3184,fumino09_8790,Fumino12_6236,Fumino15_2792}, where HB forces manifest themselves in a blue-shift with respect to the dispersion dominated band position. In the region of CH- and NH-stretching vibrations, the H-bonded vibrational modes appear red-shifted with respect to the non H-bonding modes. Furthermore, nonlinear spectroscopy has been employed to investigate the influence of H-bonding on the dephasing dynamics of the CH-stretching vibrations~\cite{Chatzipapadopoulos15_2519}.

\begin{figure}[t]
\begin{center}
\includegraphics[width=0.8\columnwidth]{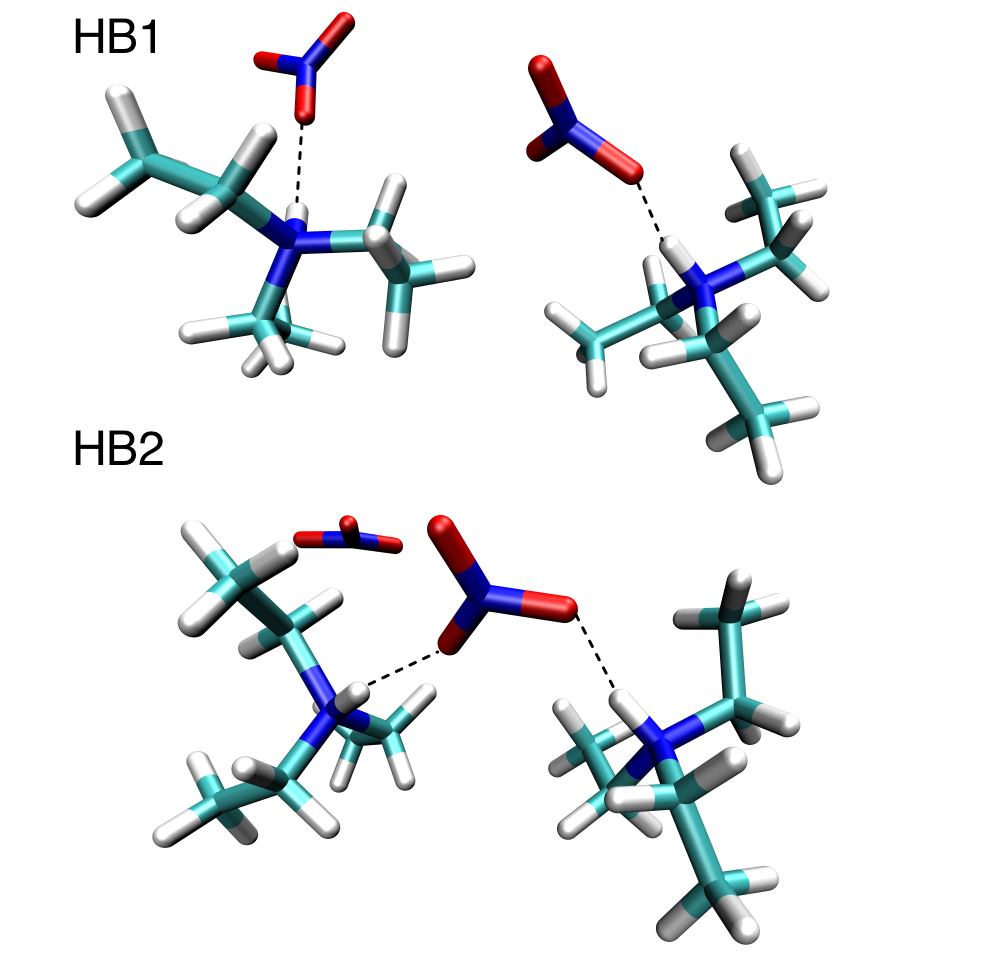}
\end{center}
\caption{Two ion pairs of triethylammonium nitrate, [C$_6$H$_{15}$]$^+$[NO$_3$]$^-$ (tEAN) with indicated HBs. The upper and lower panel show the single and double HB configuration, respectively. Correlations are quantified on the basis of the dipole moments of the NH-bonds and the ion pairs. The average absolute dipole moments for a representative case are 0.05~D (NH) and and 0.67~D (pair).}
\label{fig:motifs}
\end{figure}

Computer simulations are a useful tool to investigate properties of liquids \cite{Marx10_,Griebel07_} and various methods, ranging from coarse grained molecular dynamics (MD)~\cite{Wang07_1193} to quantum chemical calculations have been applied to ILs  (for an overview, see Refs.~\cite{kirchner09_213,Kirchner15_202}). Systems consisting of hundred thousands  of atoms are routinely simulated with  various  force fields~\cite{Dommert12_1625}. Furthermore, atomic polarization effects can be incorporated thus improving the agreement with experiments, but at a higher computational cost \cite{Schmollngruber15_14297,Schroder11_12240, Schroder12_3089}. A yet more realistic description  is provided by  ab initio MD simulations, which  were performed for a few selected small systems. For example, the ion pair and HB dynamics of methylammonium nitrate was investigated by Zahn et al. \cite{Zahn10_124506}, finding that only 1.8 out of 3 possible HB sites are H-bonded and that preferred angles with respect to other HB exist. The ion pairs are long lived, however the lifetime of the individual conformations was short, a behaviour that is called ion cage rattling. Power spectra calculated from the velocity autocorrelation functions of ab initio MD simulations can be employed to get insights into the dynamics at a microscopic level \cite{Wendler12_1570}. The experimental spectra of imidazolium based ILs could be reproduced for clusters of 8 monomers and  10~ps trajectory length, suggesting a locality in time and space that is consistent with the ion cage interpretation. 
A computationally less demanding approach is provided by the self-consistent charge density functional based tight-binding method (DFTB)  \cite{Elstner98_7260,elstner06_316,koskinen09_237}. DFTB works without empirical input and provides self-consistent Mulliken charges, which account for polarization effects. As far as ILs are concerned, it was shown, for instance, that the structure extracted from DFTB simulations matches the predicted structures of various protic ILs, including alkylammonium nitrates, from diffraction experiments \cite{Addicoat14_4633}.

\begin{figure}[th]
\includegraphics[width=0.95\columnwidth]{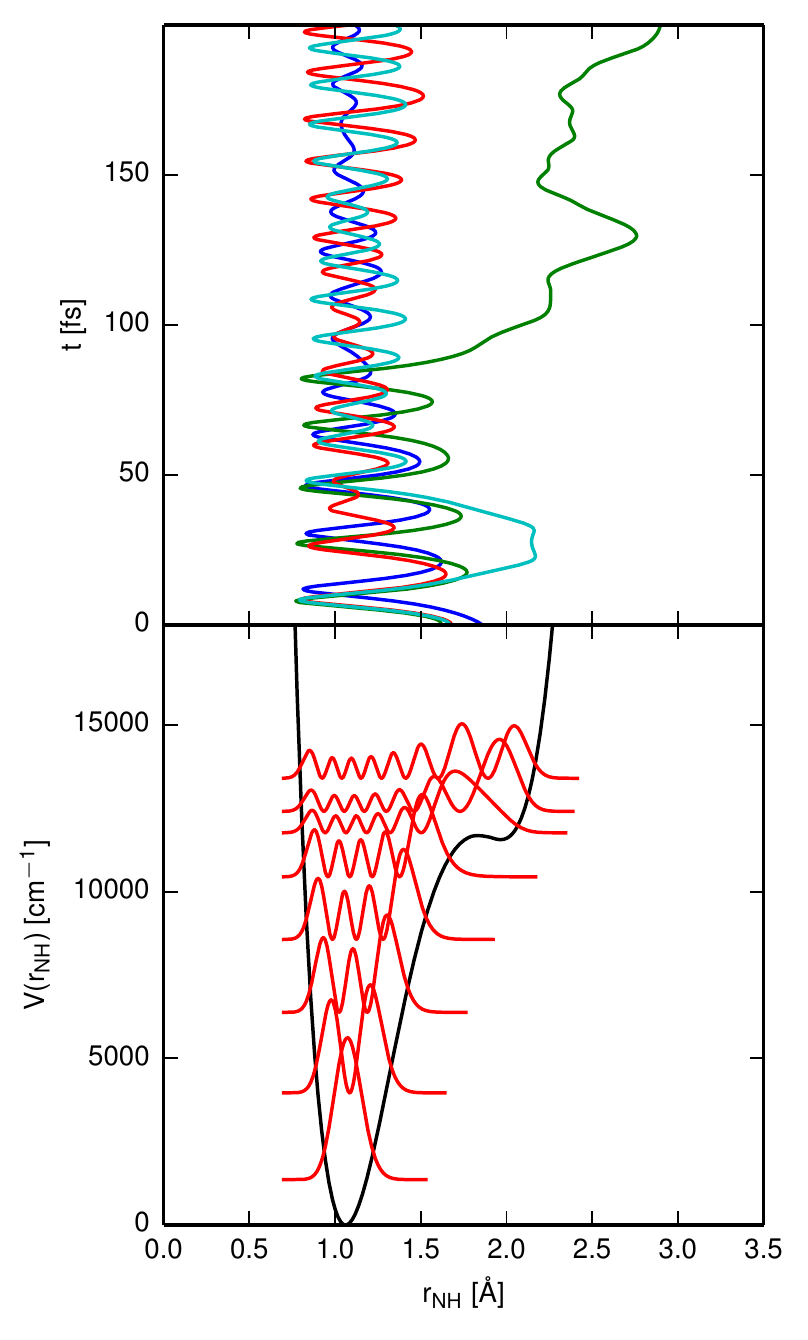}
\caption{Bottom panel: Representative snapshot of a potential curve along the bond length $r_{\mathrm{NH}}$ and respective one-dimensional vibrational states. The potential is generated by adjusting the bond lengths $r_{\rm NH} $ between $ 0.7~\mathrm{\AA}$  and $2.4~\mathrm{\AA}$ with all other atoms being fixed. Top panel: Molecular dynamics trajectories during the first 200~fs after the excitation for four typical realizations of the nonequilibrium setup.}
\label{fig:potdyn}
\end{figure}

%

In the present contribution we will use DFTB to explore the dynamics of protic ILs.  As a particular example, triethylammoniun nitrate (tEAN) will be used, where each isolated ion pair is capable of forming a single {N-H$\cdots$O} HB, cf. Fig.~\ref{fig:motifs} (upper panel). The focus will be on the quantification of correlated motions caused by intermolecular interactions. To this end, ideas from network theory are introduced~\cite{Boccaletti06_175,Brandes05_}. The ion pairs will be considered as nodes of a network graph, which are connected by edges. The correlation between the pairs is quantified by the weight of the respective edge. These weights are introduced such as to relate to different types of motions, from the proton stretching vibration to the ion pair's hindered rotation. A particular role is played by disruption and formation of ion pairs. Two situations will be considered. First, the equilibrium case, i.e.\ the system performs only thermal fluctuations within a canonical ensemble at a given temperature. Second, a proton transfer reaction is triggered by artificially raising the  energy of the proton within the HB. This way we mimic infrared excitation, whose conditions are such as to promote proton transfer and thus formation of a pair of neutral molecules. This situation corresponds to a rather drastic disturbance of the system and correlations might appear in the subsequent relaxation back to equilibrium.

The paper is organized as follows. In Section \ref{sec:methods} we start with a discussion of the MD simulation setup. Next, we provide some details concerning the generation of nonequilibrium initial conditions. We conclude this section by introducing the correlation measures used to analyse the results. Simulation results are presented in Section \ref{sec:results} for both, equilibrium and nonequilibrium cases. Finally, a summary is provided in Section \ref{sec:summary}.

\section{Theoretical Methods}
\label{sec:methods}
\subsection{Molecular Dynamics}

Molecular dynamics simulations are performed for a  box consisting of 32 tEAN ion pairs with periodic boundary conditions.  To simulate the liquid phase the temperature is set to $177^\circ \mathrm{C} $, well above the melting point $113-114^\circ$C \cite{greaves08_206}, at a density of $\rho_{\mathrm{tEAN}} = 1.048 \mathrm{g/cm^3}$.  To obtain equilibrated starting structures first classical MD is performed, which is followed by production runs with the DFTB method. For equilibration the box is simulated with Gromacs 4.5.5 \cite{hess08_435} with force field parameters taken from OPLS-AA \cite{rizzo99_4827} and its extension to ILs (especially partial charges were recalculated) by  P\'adua and coworkers \cite{canongialopes04_16893,canongialopes04_2038}. Long-range Coulomb forces were evaluated using the particle mesh Ewald summation, long range cut-off radii of 0.9 nm and a timestep of 0.5 fs are used. 
 
Production run DFTB simulations were done with the DFTB+ code \cite{aradi07_5678} including the  3rd order correction \cite{yang07_10861} and  employing the corresponding Slater-Koster parameters 3ob \cite{gaus13_338}. Van der Waals dispersion forces are included a posteriori in the DFTB calulations, with parameters taken from the Universal Force Field \cite{seifert12_456}. All runs were performed at the $\Gamma$-point only. The equilibrated structure obtained from the force field was used to start a canonical ensemble simulation in DFTB. From the respective trajectory  five starting structures for a microcanonical trajectory were sampled randomly. These DFTB production runs were simulated up to a length of 25 ps using a timestep of 0.5 fs. The microcanonical ensemble was chosen in order to have the system free from external perturbations. This way the observed response of the system to the excitation of the NH-bond is only due to the perturbation of the bond and not from the random forces of a thermostat or barostat. 

Each of the five starting structures is run in the equilibrium and nonequilibrium setup. In the nonequilibrium setup a single selected NH-bond in the box is excited by repositioning the respective proton as described below in Section \ref{sec:excitation}. All other positions and momenta are equal to the initial conditions of the equilibrium setup.  In the simulation box, on average about  18 \% of the HBs motifs have an anion, which is H-bonded to two cations (cf. Fig.~\ref{fig:motifs} (lower panel)). This suggests that for these cases different type of correlations between ion pairs may be present.  To account for these particular structural motifs, we have selected two nonequilibrium starting geometries from each of the equilibrium trajectory snapshots. These are the cases with a single and with a double HB, which yields three 25~ps  trajectories for each starting geometry to be used for the correlation analysis. The configurations with a single HB between anion and cation will be called HB1, whereas those having an anion H-bonded to two cations are will be labeled as HB2.
\begin{figure}[htb]
  \centering
  \includegraphics[width=0.95\columnwidth]{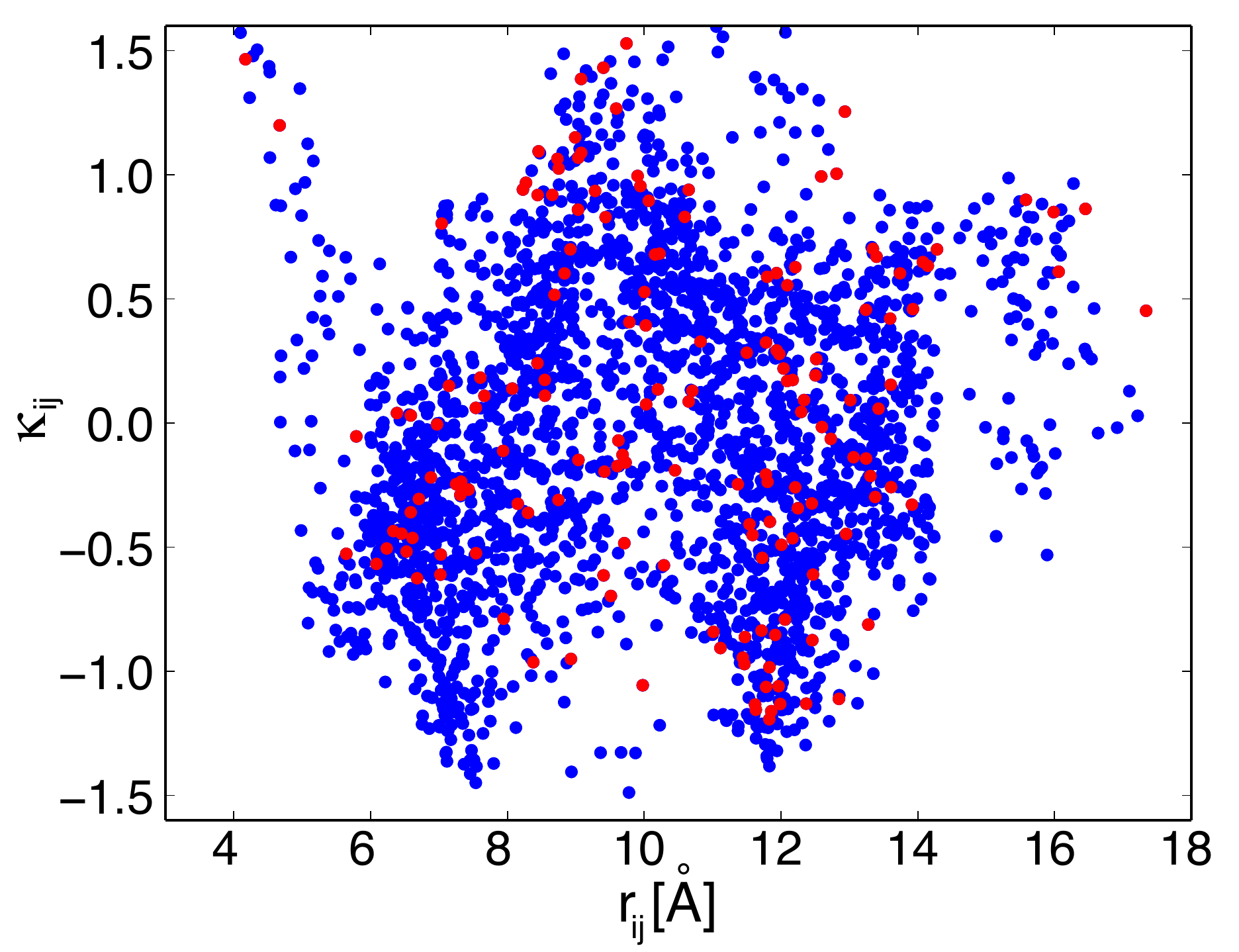}
  \caption{Distribution of average distances and relative orientations within the simulation box. Red dots mark those cases which are covered by the nonequilibrium simulation.}
  \label{fig:rkappa}
\end{figure}

\subsection{Modeling the Local Excitation}
\label{sec:excitation}
A typical potential energy curve for the displacement of the proton in an {N-H$\cdots$O} HB is shown in Fig.~\ref{fig:potdyn} (bottom panel), where it was assumed that all other atoms are fixed. The solutions of the vibrational Schr\"odinger equation for this potential are shown as well. They have been obtained using the standard Fourier Grid Hamiltonian method \cite{Marston89_3571}. To have the possibility of a transfer and to create a  pair of neutral molecules, the probability density of the vibrationally  excited state must be reasonably high at the ’second’ minimum. In the representative potential plotted in Fig.~\ref{fig:potdyn} this was only the case for $\nu > 5$. Hence,  in the following excitation to the $\nu_{\mathrm{excited}}=6$ level will be assumed to generate a nonequilibrium initial condition.  To model this excitation of the bond for the general situation, the proton is moved in $0.03~\mathrm{\AA}$  steps along the NH-bond vector and the potential energy of the new positions is calculated. Starting from the position where the  potential energy is closest to the energy of the of the vibrational level $\nu_{\mathrm{excited}}$, the procedure is repeated with a step size of 0.01~$\mathrm{\AA}$. As soon as the  energy approaches the energy of the vibrational level $\nu_{\mathrm{excited}}$, the respective geometry is used to start the nonequilibrium trajectory. The kinetic energy remains unchanged.

\subsection{Correlation Coefficients}
In general one expects that due to the strong interaction with the nitrate anion all cations form HBs, and the IL can be described as consisting of ion pairs. The pair dipole, pointing roughly from the cationic nitrogen to the anion center, will be a reasonable quantity for the analysis of  correlated motions of different pairs. The HB motion itself might be different from the more global ion pair dynamics and a more specific measure is needed. Here, we will use the NH-bond dipole, which would be observable in infrared spectroscopy.
Thus, we have two types of dipole moments, $\vec{d}$, for quantifying correlations, which might be present in different types of motion and on different time scales.

The present analysis of the equilibrium trajectories provides evidence that the H-bonding between ion pairs is weak enough such as to allow frequent changes of the H-bonding partners, including a switch from an HB1 to a HB2 configuration. In fact such changes occur on average every 1.6~ps. 
They represent an abrupt change with respect to the above dipole moments. As a  consequence, the dipole moments are not normally distributed on the time scale of the trajectory and hence the often used Pearson correlation coefficient analysis is not applicable. Instead, in order to quantify the correlations of the dipole moment  dynamics, the non-parametric Spearman's rank correlation coefficients $r_{\rm S}$ are used. Given two  time series $x_t$ and $y_t$ with the time index $t$, Spearman's rank correlation coefficient is defined as \cite{Spearman04_72}
\begin{equation}
\!\!\!\!\!\!\!\!\!\! r_{\rm S}(x,y) \! = \! \frac{\sum_{t}(\mathcal{R}_{x_t}-\overline{\mathcal{R}}_x)(\mathcal{R}_{y_t}-\overline{\mathcal{R}}_y)} 
 {[\sum_{t}(\mathcal{R}_{x_t}-\overline{\mathcal{R}}_x)^2 \sum_{t}(\mathcal{R}_{y_t}-\overline{\mathcal{R}}_y)^2]^{1/2}}\, . 
\end{equation}
The rank  $\mathcal{R}_{x_t}$ returns the position of the value when the time series data is ordered according to the absolute values.
 $\overline{\mathcal{R}}_x$ denotes the average rank of $x$. Values of $r_{\rm S}$ can range from -1 (negative correlation) to 1 (positive correlation), with 0 being no correlation.

 Correlations will be quantified by representing the IL as a network. Here, the ion pairs take the role of the network nodes. The weight $w_{ij}$ between nodes $i$ and $j$ is calculated from the trajectories of dipole moment vectors. However, due to the abrupt changes the commonly used analysis in terms of fluctuations with respect to the mean value doesn't provide the insight we are interested in. It turned out that in order to trace the correlations between the motions of different ion pairs and their HBs, which is actually strongly influenced by the abrupt changes of bonding partners, it is useful to look at the time series of \textit{increments } of the dipole moment vector, $\Delta \vec{d}_t=\vec{d}_t-\vec{d}_{t-\delta t}$ ($\delta t$: MD step size), of the ion pairs and the NH-bond distance. This gives the respective edge weights, $w_{ij}$, which are calculated from the correlation coefficients $r_{\rm S}$ by considering correlations between all three vector components of the considered dipole moments, $\Delta \vec{d}=(\Delta d_1,\Delta d_2,\Delta d_3)$. Averaging over these components gives
\begin{equation}
w_{ij} = \frac{1}{9}\sum_{m,n=1}^3 |r_{\rm S}(\Delta d_n,\Delta d'_m)|   \, , 
\label{eq:weight}
\end{equation}
for two nodes $i$ and $j$ characterized by the dipole increments $\Delta\vec{d}$ and $\Delta\vec{d}'$, respectively The weights are normalized such that they have values between 0 and 1.  Note that it has been checked that the weights are converged with respect to the length of the trajectory.

\begin{figure}[htb]
  \centering
  \includegraphics[width=0.8\columnwidth]{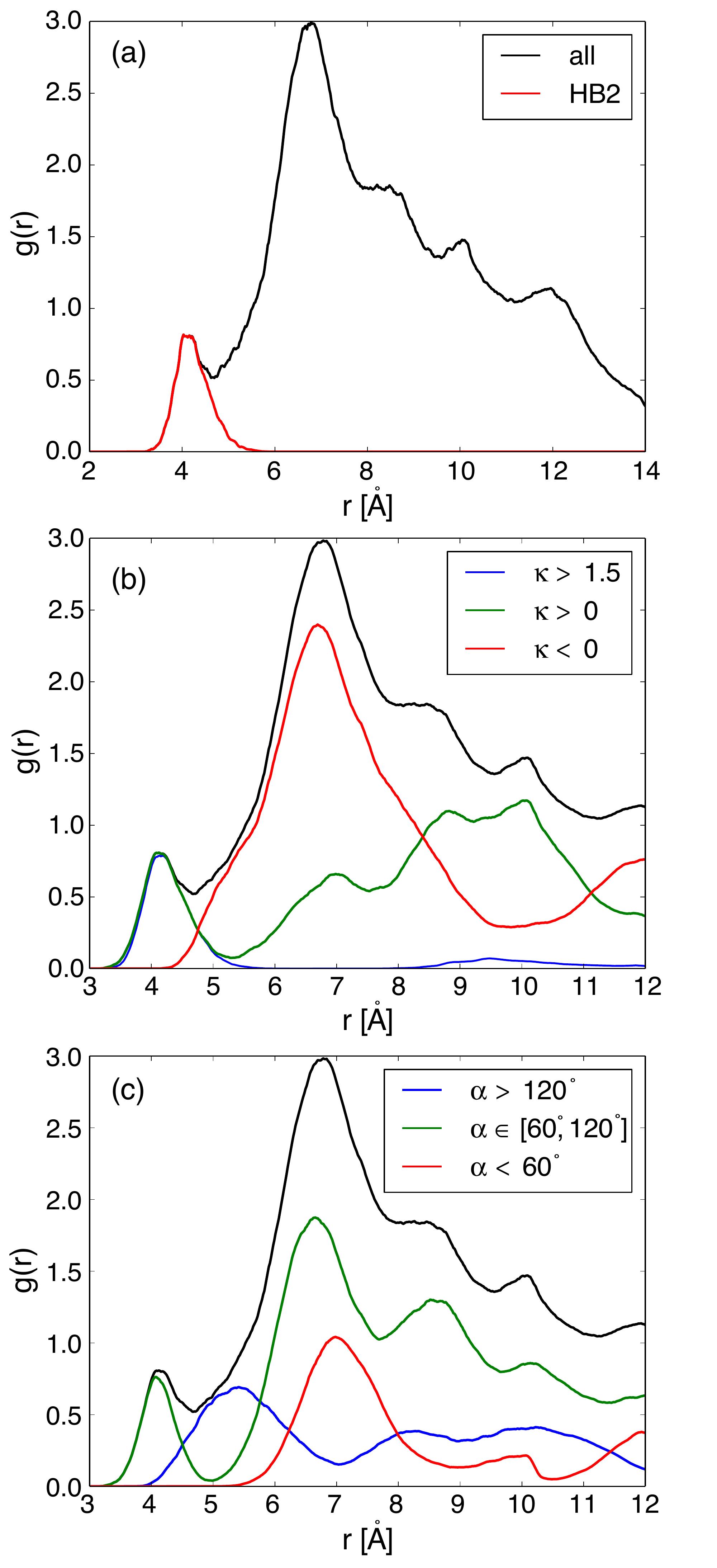}
  \caption{(a) Radial distribution function, $g(r)$, with respect to the center of mass coordinates defining the dipole moments (black line) and contribution of HB2 motifs (red line). (b) Decomposition of $g(r)$ according to the contributions from different ranges of the orientation factor $\kappa$. (c) Decomposition of $g(r)$ according to the contributions from different angles, $\alpha$, between the two pair dipole moments.}
  \label{fig:gr}
\end{figure}

\section{Results}
\label{sec:results}

\subsection{Structural Analysis}
\label{sec:structural-analysis}

In this section we will consider the IL as a network consisting of nodes and edges. The nodes are positioned at the center of mass of the ion pair and the edges connect all the nodes. Each edge carries a weight calculated according to Eq.~(\ref{eq:weight}), that represents the correlation strength between the two nodes. First, we take a look at the unperturbed (equilibrium) network and analyze the correlations between the dipole moments of ion pairs and NH-bonds. Subsequently, equilibrium and nonequilibrium cases will be compared. In order to link correlations to local structure, the edge weights will be analyzed with respect to the distance between the nodes, $r_{ij}$, and the  relative orientation of the dipoles associated with the nodes. For the latter we use the orientation factor
\begin{equation}
	\kappa_{ij} =  \vec{\underline{d}}_i \vec{\underline{d}}_j - 3 (\vec{\underline{r}}_{ij} \vec{\underline{d}}_i) (\vec{\underline{r}}_{ij} \vec{\underline{d}}_j)
\end{equation}
where $\vec{\underline{d}}=\vec{d}/|\vec{d}|$ and $\vec{\underline{r}}=\vec{r}/|\vec{r}|$ denote the  unit vectors. 
%

The distribution of \emph{average} distances and relative orientations in the simulation box is shown in Fig.~\ref{fig:rkappa}. Noteworthy in this distribution is the ``band'' of short distances and $\kappa_{ij}$-values beyond 0.0. 

Anticipating that correlations will be of particular importance at short distances, the structure of the IL is analyzed in terms of the radial distribution function, $g(r)$, with respect to the center of mass coordinates of the ion pairs. 
In Fig.~\ref{fig:gr}a we show $g(r)$ for all pairs (black) and for HB2 motifs only (red). Apparently, the first peak at short distances (around 4~\AA) is solely due to HB2 motifs. Panel b of Fig.~\ref{fig:gr} shows the decomposition of the radial distribution function with respect to different ranges of $\kappa_{ij}$. It is nicely seen that the HB2 motif comes with orientation factors  $\kappa_{ij}>1.5$. Finally, in  Fig.~\ref{fig:gr}c a decomposition in terms of angles between the pair dipole moment vectors is provided. The HB2 motif which dominates the peak around 4~\AA{} facilitates angles between 60$^\circ$ and 120$^\circ$.  Preparing for the following discussion it should be stressed again that HB1 and HB2 motifs interconvert on average every 1.6~ps. Thus along a trajectory an ion pair will contribute to the HB2 peak around 4~\AA, but also to the range of the maximum around 6.5~\AA{}. In passing we note that while being in the HB2 configuration,  the notion of two pair dipoles is, of course, a bit misleading since the two pairs share the same anion. 
\subsection{Equilibrium Case}
The edge weights representing correlations between the motions of the different dipole moments  are relatively small, if averaged with respect to all pairs in the simulation box. Specifically, we have for the pair and NH dipoles 0.038 and 0.040, respectively.
\begin{figure}[tbh]
  \centering
  \includegraphics[width=0.8\columnwidth]{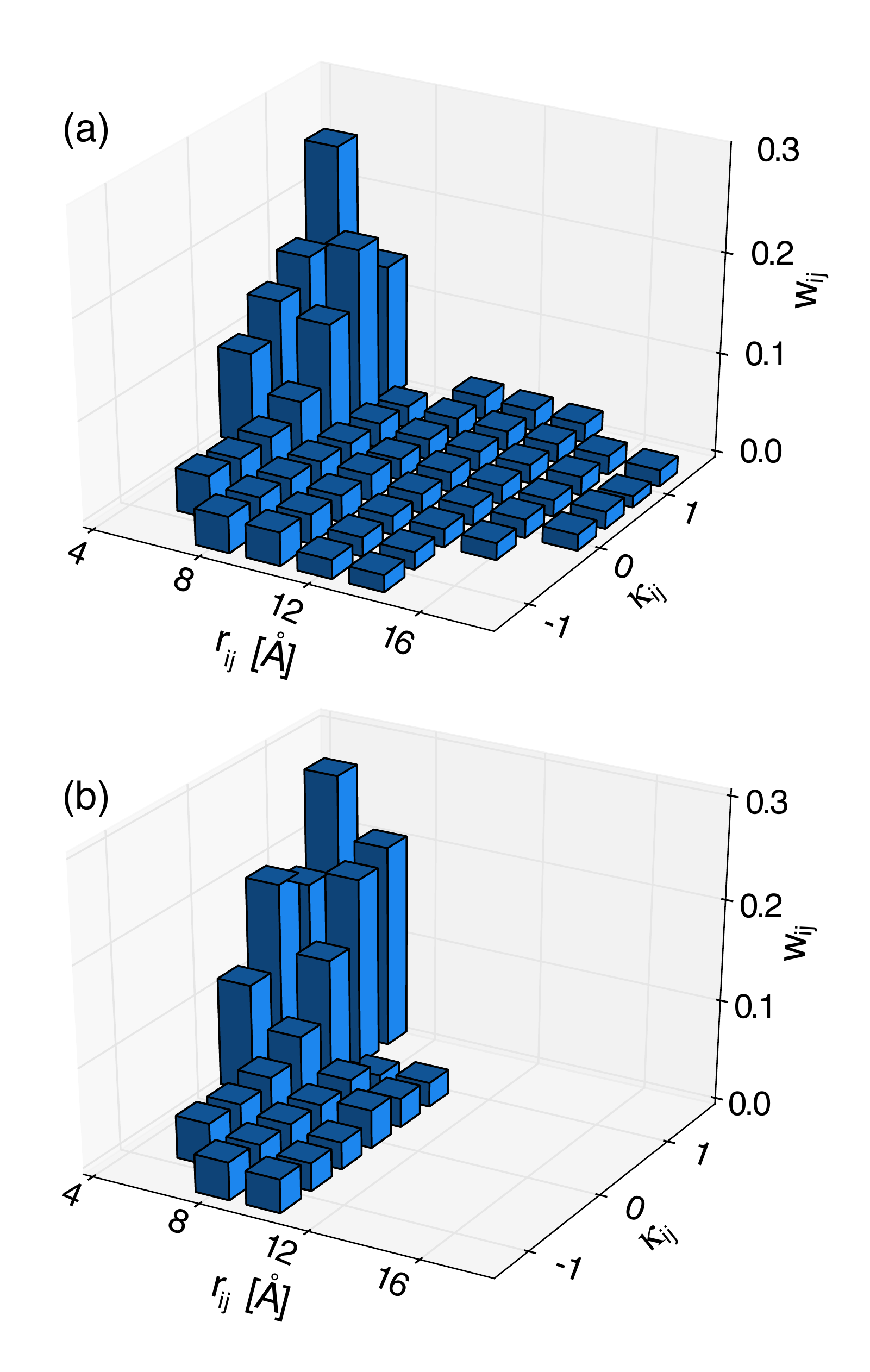}
  \caption{Dependence of the correlation edge weights on the distances and orientations between the pair dipole moments for all pairs (a) and for those having an HB2 motif (b).}
  \label{fig:pairdip}
\end{figure}
A more detailed view on the correlated motion is provided by the distance and relative orientation dependence of the edge weights.In contrast to $g(r)$ the edge weights $w_{ij}$ are plotted as a function of position and orientation factor averaged over the whole trajectory. In Fig.~\ref{fig:pairdip}a the case of pair dipole moments is shown. Clearly, correlations play a role for short distances only, i.e. up to the maximum in $g(r)$ in Fig.~\ref{fig:gr}. Further, there is a preference for those configurations, which have parallel dipoles ($\kappa_{ij}\approx 1$). In  Fig.~\ref{fig:pairdip}b the edge weights are plotted for those pairs that have an HB2 motif at some point along the trajectory. Comparing both panels and Fig.~\ref{fig:gr} one can conclude that correlated motion between ion pairs takes place mostly for the HB2 configuration. Or, in other words, the longer a pair stays in a HB2 configuration along the trajectory the stronger is the correlation as measured by the edge weight.
A similar analysis has been performed for the NH-bond dipoles (see Suppl. Mat., Fig. 1). In this case, however, we find no distance and orientation dependence.
\subsection{Nonequilibrium Case}
\label{sec:nonequilibrium-case}

\subsubsection{Vibrational Energy Redistribution}
Before analysing the fluctuation dynamics in detail, the vibrational energy redistribution following nonequilibrium excitation of an NH-bond will be investigated. To this end, all 32 ion pairs in each of the five starting geometries are prepared and nonequilibrium runs for 200~fs are performed. Typical trajectories are shown in Fig.~\ref{fig:potdyn} (top panel). After 'excitation' the proton motion has a large amplitude, but the excess energy is rapidly dissipated into the environment. Eventually, the proton gets trapped either in its original configuration (ion pair) or transiently at the HB acceptor side (two neutral molecules). Out of the 160 nonequilibrium setups only one pair of neutral molecules is present after 200~fs (cf. green line in Fig.~\ref{fig:potdyn} (top panel)). Hence, proton transfer is a rather unlikely event and the ionic configurations are strongly favoured due to the ionic background \cite{hunt15_1257}.   The average N-H distance of the modified cation at the beginning of the run was $r_{\mathrm{NH}} =  1.74~\mathrm{\AA} $. After 200~fs the average bond distance of the excited cations is $r_{\mathrm{NH}} =  1.08~\mathrm{\AA}$, i.e.\ close to the equilibrium bond distance of  $r_{\mathrm{NH}}  = 1.05~\mathrm{\AA}$.

\begin{figure}[h!]
\includegraphics[width=\columnwidth]{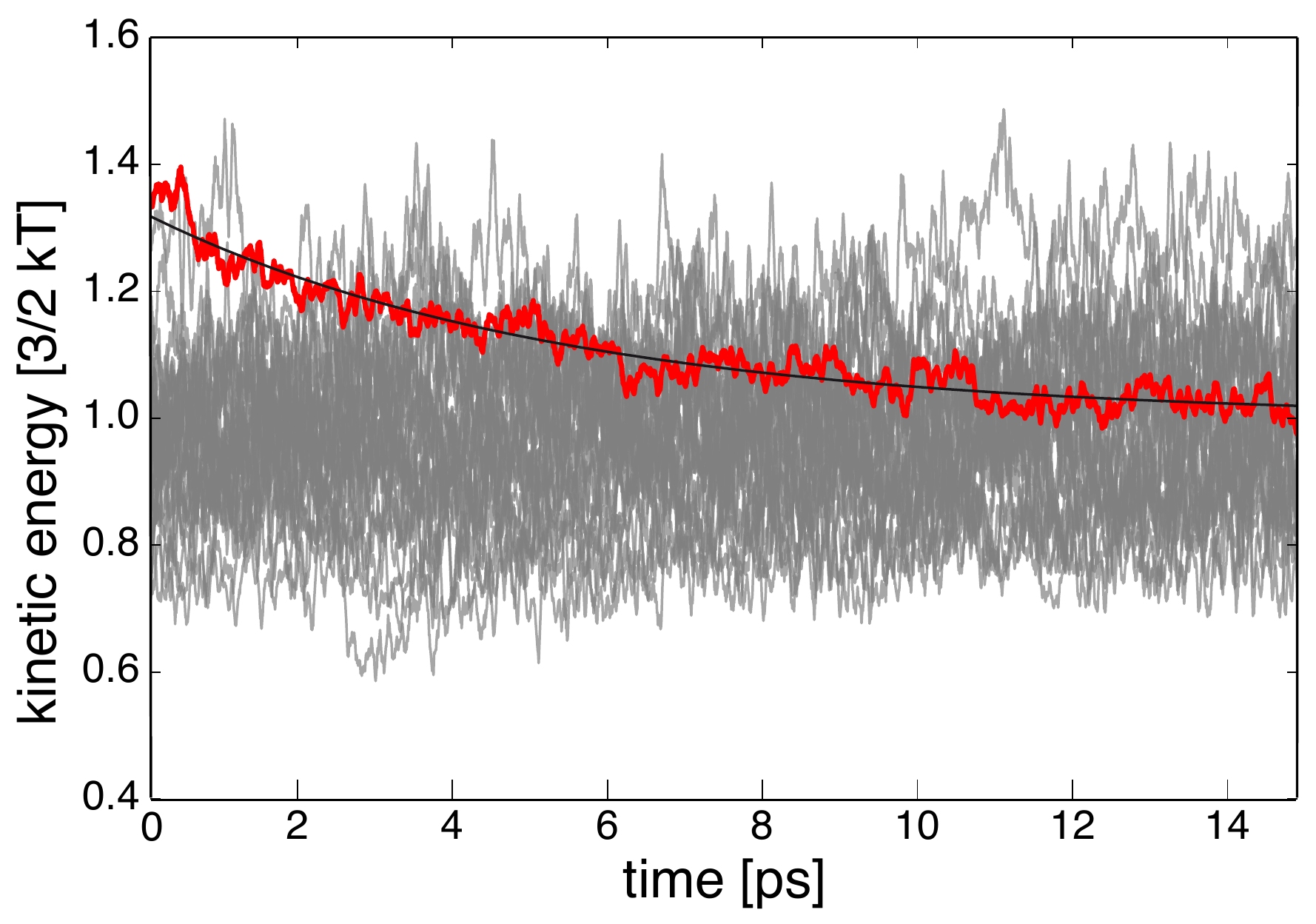}
\caption{Kinetic energy of the excited cation (red) averaged over 10 trajectories and its monoexponential fit with time constant of 5.3~ps. The grey lines show the kinetic energy of  all other 31 other cations for one selected trajectory.}
\label{fig:Ekin}
\end{figure}

 The kinetic energy of the cations is calculated as the sum of the atomic kinetic energies and the moving average of 50~fs is plotted in Fig.~\ref{fig:Ekin} for an exemplary case.  By fitting the kinetic energy of the excited cation to   $ \langle E_{\mathrm{kin}}\rangle (t) = E_{\mathrm{excited}}\times\exp(-t/\tau)$, a relaxation time $\tau$ can be obtained. In the five 25~ps NVE runs the average relaxation time  is $\tau = 5.3$~ps. Thereby, the relaxation time of cations which form a HB to anions that themselves are not involved in another HB, $\tau_{\mathrm{HB1}} = 7.0$~ps, is larger than the relaxation time $\tau_{\mathrm{HB2}} = 4.1$~ps of cations H-bonded to anions that form another HB. The extra HB on the anion appears to trigger extra dissipation pathways. However,  the intuitive picture does not apply, i.e. neither the  counter anion nor the second cation H-bonded to the anion do show a noticeable increase in kinetic energy (not shown).   In summary, the average relaxation time is 5.3~ps, but specific values depend on the local structure of the HBs.
 
\subsubsection{Correlation Analysis}
\begin{figure}[t]
  \centering
  \includegraphics[width=\columnwidth]{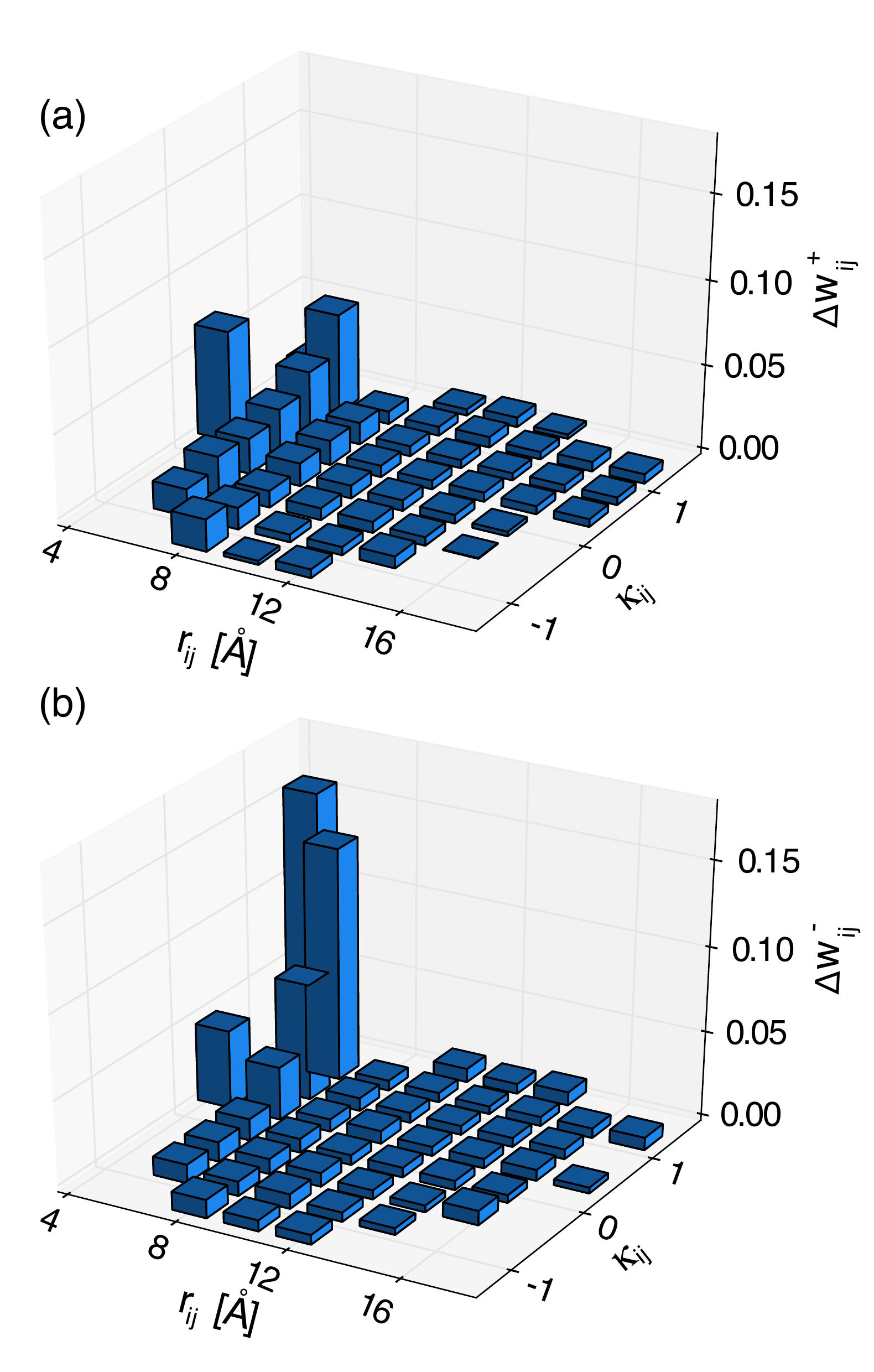}
\caption{Change of the pair dipole correlation edge weights between nonequilibrium and equilibrium simulations, $\Delta w_{ij} = w_{ij}^{\rm (non)} - w_{ij}^{\rm (eq)}$, as a function of the distances and orientations between the pair dipole moments. The upper (a) and lower (b)  panel show the absolute value of the positive ($\Delta w_{ij}^+$) and negative ($\Delta w_{ij}^-$) change, respectively.}
\label{fig:wchange}
\end{figure}

The excitation of the selected ion pair perturbs the dynamics of all ions and therefore influences the correlation between the pairs. According to the analysis of the previous section, the energy disposed into the system is rapidly redistributed without any preference for a certain pathway.  In Fig.~\ref{fig:wchange} we give the distance and orientation resolved change of the pair dipole correlation edge weights between nonequilibrium and equilibrium simulations, $\Delta w_{ij} = w_{ij}^{\rm (non)} - w_{ij}^{\rm (eq)}$. Overall,  changes are rather local, i.e. within the range where correlations have been present already in the equilibrium simulation. Interestingly, the largest changes are observed for those distances and orientations where the correlations in equilibrium have been strongest, but, here $\Delta w_{ij}<0$, i.e. the edge weights decrease. It is tempting to assume that this is an effect of local heating, which increases the fluctuations such that correlated motion between neighboring ion pairs is disturbed.

\begin{figure}[bth]
\includegraphics[width=\columnwidth]{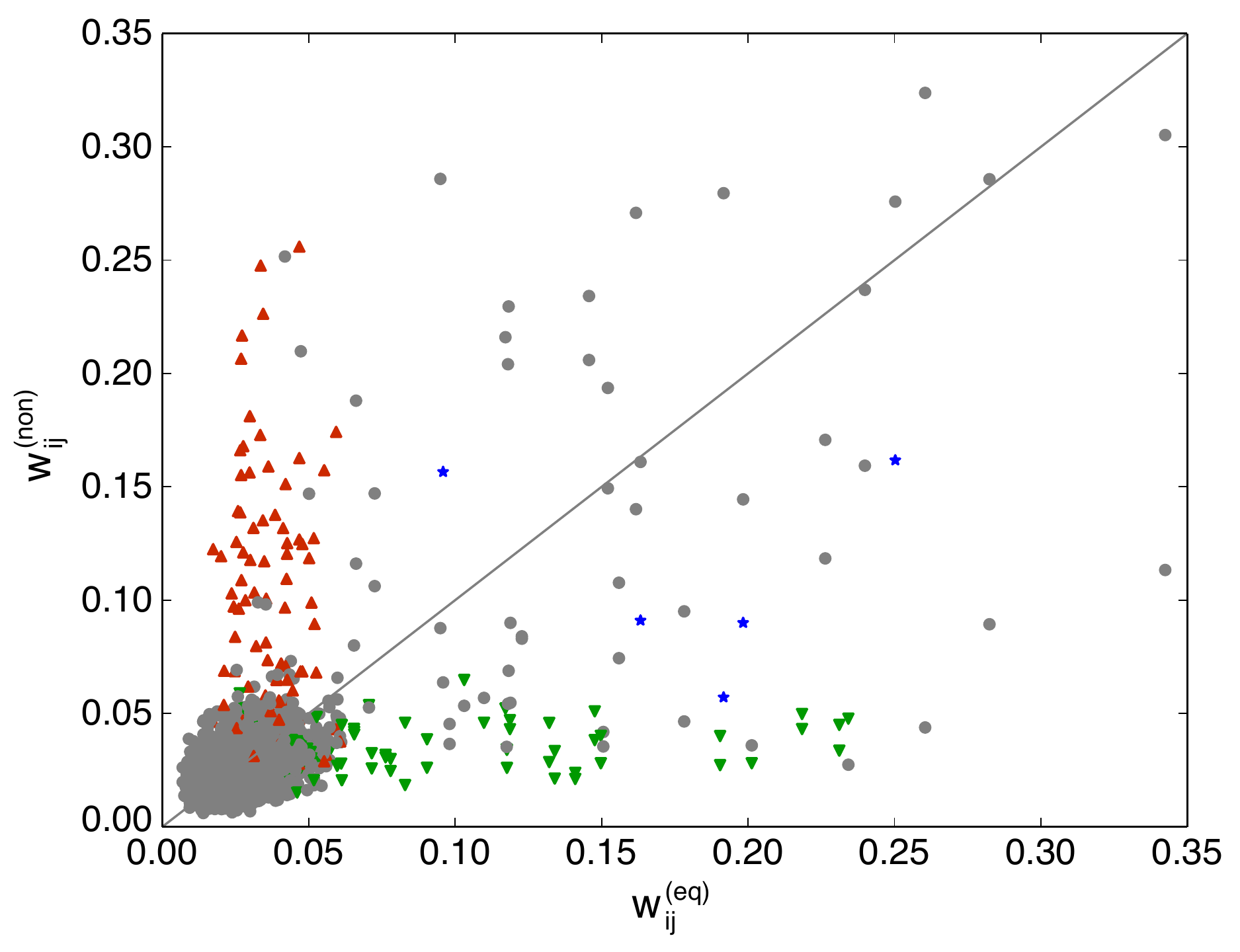}
\caption{Edge weights of pair dipoles for the equilibrium vs. the nonequilibrium simulation. The blue stars denote those pairs that have been initially excited in the nonequilibrium simulation. Green symbols denote those pairs that are in a HB2 motif in the equilibrium but not in the nonequilibrium simulation, whereas red symbols denote those pairs where the opposite is true. The bisecting line is drawn to guide the eye.}
\label{fig:weqvsqnon}
\end{figure}

A more detailed picture is provided in Fig.~\ref{fig:weqvsqnon}, where the edge weights of all pair dipoles are shown for equilibrium $w_{ij}^{\rm (eq)}$  vs. nonequilibrium $w_{ij}^{\mathrm{(non)}}$  simulations. Apparently, there is no linear correlation between $w_{ij}^{\rm (eq)}$ and $w_{ij}^{\rm (non)}$, which implies that the pattern of ion pair correlations changes completely. Having in mind the special role of HB2 configurations, the respective pairs have been highlighted for equilibrium and nonequilibrium simulations. Here, a pair is counted as HB2 when it persists for more than 500~fs along the trajectory. Inspecting Fig. ~\ref{fig:weqvsqnon} we note that there are pairs, which have a large  $w_{ij}^{\rm (eq)}$ but a small $w_{ij}^{\mathrm{(non)}}$ and the other way around. For the former it holds that these pairs where in a HB2 configuration during the equilibrium simulation, but turned into a HB1 configuration in the nonequilibrium case. For the latter pairs the situation is opposite. Closer inspection of the trajectory revealed that the initial excitation spreads over the simulation box, which causes HB1 into HB2 conversions and vice versa. Nevertheless, this has no effect on the local nature of the important edge weights.  However, these weights are calculated with respect to the total trajectory, i.e. transient correlations might not be properly captured.

\section{Summary}
\label{sec:summary}
Due to their ability to form HBs, protic ILs are often considered as networks~\cite{fumino09_3184,fumino09_8790}. For the simple case of triethylammonium nitrate, we have addressed the question to which extent such networks actually exist. Taking a dynamics perspective a network should imply correlated motions of ion pairs. Using ideas from network theory, the ion pairs were considered as nodes, connected by edges. The weight of a certain edge was determined from the Spearman correlation coefficient, which was calculated on the basis of the fluctuating dipole moments.

For the present example it was found that correlations are on average weak. However, sizeable local correlations between the pair dipole moments exist. Here, structural motifs play a prominent role, where two cations share one anion with two {N-H$\cdots$O} HBs. Correlations between the dipole moments of the NH bonds are of minor relevance. A local nonequilibrium situation introduced  by artificially exciting a single NH-bond vibration rapidly equilibrates on a time scale of 5.3~ps. This changes the distribution of correlated pairs in the simulation box, but does not lead, e.g., to long-range correlations spreading from the initially excited pair.

Finally, we address the question how to observe correlated motions in protic ILs. In principle, two-dimensional infrared spectroscopy is the method of choice \cite{hamm11_}. In the present case, the potential target could be the NH-stretching vibration, which is part of the HB. Correlated or anti-correlated NH-vibrations would show up in specific off-diagonal features of the two-dimensional spectrum \cite{yan11_5254}.
Based on the present results, triethylammonium nitrate cannot be recommended for such a study. However, it can be anticipated that stronger correlations exist for those cases where in the  triethylammonium cation one or both alkyl chains have been replaced by hydrogen, to facilitate additional H-bonding.
\section*{Acknowledgements}
The authors thank the Deutsche Forschungsgemeinschaft (DFG) for financial support through the SFB 652.

\bibliography{IL_tEAN_2}

\end{document}